\documentclass[a4paper,pra,twocolumn,superscriptaddress]{revtex4-2}

\usepackage{amssymb}
\usepackage{amsmath}
\usepackage{epsfig}
\usepackage{color}
\usepackage{graphics, graphicx}
\usepackage{bbold}
\usepackage{psfrag}
\usepackage{mathcomp}
\usepackage{subfigure}
\usepackage{verbatim}
\usepackage{color}
\usepackage{blindtext}
\usepackage{float}
\usepackage[utf8]{inputenc}
\usepackage[T1]{fontenc}
\usepackage[colorlinks,citecolor=blue]{hyperref}

\begin{document}

\date{\today}
\title{{Superradiant} phase transitions in one-dimensional correlated Fermi gases with cavity-induced umklapp scattering}

\author{Jian-Song Pan}
\email{panjsong@scu.edu.cn}
\affiliation{College of Physics, Sichuan University, Chengdu 610065, China}
\affiliation{Key Laboratory of High Energy Density Physics and Technology of Ministry of Education, Sichuan University, Chengdu 610065, China}

\begin{abstract}
The {superradiant} phase transitions of one dimensional correlated Fermi gases in a transversely driven optical cavity, under the umklapp condition that the cavity wavenumber equals two times of Fermi wavenumber, are studied with the bosonization and renormalization group (RG) techniques. The bosonization of Fermi fields gives rise to an all-to-all sine-Gordon (SG) model due to the cavity-assisted non-local interactions, where the Bose fields at any two spatial points are coupled. {The superradiant phase transition is then mapped to the Kosterlitz-Thouless phase transition of the all-to-all SG model. The nesting effect, in which the superradiant phase transition can be triggered by infinitely small atom-cavity coupling strength, is shown to be preserved for any non-attractive local interactions. For attractive local interactions, the phase transition occurs at finite critical coupling strength. Nevertheless, the analysis of scaling dimension indicates that the perturbation of non-local cosine term is indeed relevant (irrelevant) when the scaling dimension is lower (higher) than the critical dimension, similar to the case of ordinary local SG model. Our work provides an analytical framework for understanding the superradiant phase transitions in low-dimensional correlated intra-cavity Fermi gases.}
\end{abstract}
\pacs{67.85.Lm, 03.75.Ss, 05.30.Fk}

\maketitle
\section{Introduction}
The quantum gases in an optical cavity provide new paradigms for exploring many-body physics~\cite{ritsch2013cold,mivehvar2021cavity}. The self-organization of atoms into a checkerboard order, accompanying the superradiant macroscopic occupation of cavity modes, occurs above a critical atom-cavity coupling strength~\cite{baumann2010dicke,klinder2015observation}. For Bose-Einstein condensate (BEC), the superradiant phase transition is approximately characterized by a generalized Dicke model since atoms only condensate into the several resonant modes~\cite{baumann2010dicke}. The superradiant phase transition of Fermi gases (and hard-core bosons~\cite{rylands2020photon}) shows distinct features due to the presence of Fermi surface~\cite{piazza2014umklapp,keeling2014fermionic,chen2014superradiance,pan2015topological,mivehvar2017superradiant,yu2018topological,zhang2021observation}. {For example, when the Fermi surface is commensurate with the cavity wave length $k_{0}=2k_{F}$ ($k_{F}$ is the Fermi wave number), i.e., the atoms are scattered by cavity photons with the umklapp condition, the nesting effect that the critical point softens to zero is predicted in one dimensional Fermi gases~\cite{piazza2014umklapp,keeling2014fermionic,chen2014superradiance}. For higher dimensional Fermi gases, the critical coupling strengthes are also prominently decreased approaching the umklapp condition in certain directions.}
Recently, the superradiant phase transition of Fermi gases was observed experimentally~\cite{zhang2021observation}.

The cavity-induced spontaneous symmetry breaking in intra-cavity quantum gases in general can be interpreted with the effective atom-atom interaction generated by the adiabatical elimination of cavity dynamics~\cite{habibian2013bose,gopalakrishnan2009emergent,leonard2017supersolid,mivehvar2019emergent}, upon the picture of cavity-induced dynamical potentials~\cite{mivehvar2014synthetic,padhi2014spin,dong2014cavity,deng2014bose,pan2015topological,kroeze2019dynamical,zheng2016superradiance,kollath2016ultracold,sheikhan2016cavity,gulacsi2015floquet,
zheng2016superradiance,ballantine2017meissner}. Although the fluctuations of single-mode cavity field in the suprradiant phases in the thermal dynamic limit may be ignorable, due to the cavity modes essentially have no spatial dynamics~\cite{piazza2013bose}, the interplay of cavity-assisted non-local and local interactions in low-dimension correlated quantum gases is still under exploring~\cite{landig2016quantum}. The supperadiant phase transition of Fermi gases crossing Feshbach resonance shows smooth crossover between the BCS and BEC regimes~\cite{chen2015superradiant,yu2018topological}. Experimentally preparing strong correlated Fermi gases in an optical cavity calls for further study on the interplay between correlation effects and cavity-induced dynamics~\cite{roux2020strongly,roux2021cavity}.

In this paper, we theoretically study the {superradiant} phase transitions of one-dimensional (1D) correlated Fermi gases trapped in an optical cavity with the bosonization and renormalization group techniques. We focus on the nesting point where the Fermi surface has a wave length commensurate with the cavity wave length, i.e., $k_{0}=2k_{F}$. The bosonization of the Fermi fields gives rise to an all-to-all sine-Gordon model, in which the Bose fields in any distances are coupled. The superradiant phase transition is then linked to the well-known $(1+1)$-dimensional KT phase transition of SG model. By employing the perturbative renormalization group techniques, we capture the ground-state phase diagram of the Fermi gas. The nesting effect, where the model shows vanishing critical coupling strength, survives only with repulsive local interaction. The critical coupling strength become finite when the local interaction becomes attractive. {The critical behaviours are discussed. Nevertheless, we find the relevancy of the non-local cosine term also subjects to the scaling-dimension analysis; it is relevant (irrelevant) when the scaling dimension is lower (higher) than the critical dimension of KT phase transition, similar to that in an ordinary local SG model.}

In the following, we present the model and its bosonization in section II and III, respectively. Further, we perform the RG analysis in section IV, discuss the phase diagram in section V, and analyze the critical dimension in section VI. A brief summary is given in the last section.

\begin{figure}
  \centering
  \includegraphics[width=8cm]{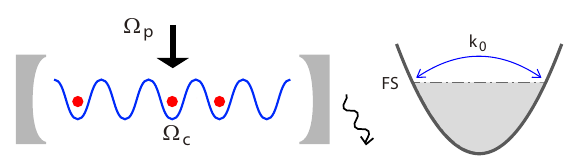}\\
  \caption{Illustration of the model. The 1D Fermi gas is coupled with the cavity field ($\Omega_{c}$) via a transverse driving ($\Omega_{p}$). The atoms are assumed to have two energy levels (not shown), and the higher level has been adiabatically eliminated since the driving and cavity fields are far detuning from the energy levels. The Fermi surface (FS) is assumed to be resonant with the cavity field, i.e., $k_{0}=2k_{f}$. }\label{Fig:model}
\end{figure}

\section{Model}
We consider an 1D Fermi gas of two-level fermions loaded in a single-mode optical cavity [see Fig.~\ref{Fig:model}], where the atomic and cavity dynamics are coupled through the scattering of a transverse driving field $\Omega_{p}$ by atoms into the cavity. The single-photon detuning $\Delta$ (i.e., the frequency differences between the energy-level difference and the light fields) is assumed to be far larger than other energy scales and then the higher level can be adiabatically eliminated. The model Hamiltonian thus is given by
\begin{equation}\label{eq:Hamiltonian}
\begin{split}
\hat{H}=&\int dx\hat{\psi}^{\dagger}(x)[-\frac{\nabla^{2}}{2m}-\mu+V_{0}\hat{a}^{\dagger}\hat{a}\cos^{2}(k_{0}x)]\hat{\psi}(x)\\
&+\eta\left(\hat{a}^{\dagger}+\hat{a}\right)\int dx\hat{n}(x)\cos\left(k_{0}x\right)-\Delta_{c}\hat{a}^{\dagger}\hat{a}\\
&+\int\int dx drU(|r|):\hat{n}(x)\hat{n}(x-r):,
\end{split}
\end{equation}
where $\hat{\psi}$ and $\hat{a}$ are the field operators for atoms and cavity photons, $\mu$ is the chemical potential of fermions, $V_{0}=\Omega_{c}^{2}/\Delta$ term with density operator $\hat{n}=\hat{\psi}^{\dagger}\hat{\psi}$ and single-photon Rabi frequency $\Omega_{c}$ represents the stark shift induced by the cavity field, $\eta=\Omega_{c}\Omega_{p}/\Delta$ with the Rabi frequency of pumping field $\Omega_{p}$ is the atom-cavity coupling strength, $\Delta_{c}$ is the two-photon detuning (i.e., the frequency difference between the cavity mode and pumping field), and the last term is the local interaction. Here the symbol $:\cdots:$ denotes the normal order. {The Fermi gas is assumed to be prepared with the umklapp condition that the cavity wavelength $k_{0}$ is commensurate with the Fermi wavelength $k_{F}$, i.e., $k_{0}=2k_{F}$. This setup is focused here mainly due to the nesting effect perfectly manifests itself in softening the critical coupling strength to zero in one-dimensional Fermi gases~\cite{piazza2014umklapp,keeling2014fermionic,chen2014superradiance}.}

The dynamics of cavity mode $\hat{a}$ can be captured by the Heisenberg-Langevin equation,
\begin{equation}
i\hbar\dot{\hat{a}}=-(\tilde{\Delta}_{c}+i\kappa)\hat{a}+\eta\int dx \hat{n}(x)\cos\left(k_{0}x\right)+\sqrt{2\kappa}\hat{a}_{in}(t),
\end{equation}
where $\tilde{\Delta}_{c}=\Delta_{c}+V_{0}\int dx \hat{\psi}^{\dagger}(x)\cos^{2}\left(k_{0}x\right)\hat{\psi}(x)$, $\kappa$ describes the decay of the cavity and $\hat{a}_{in}(t)$ satisfying $\langle\hat{a}_{in}\rangle=0$ and $\langle\hat{a}_{in}(t)\hat{a}_{in}^{\dagger}(t^{'})\rangle=\delta (t-t^{'})$ is the Langevin operator~\cite{carmichael2009open,habibian2011quantum,habibian2013bose,sieberer2016keldysh}. We assume the optical cavity has a far shorter characterisitc time $\delta t_{c}$ due to the large $\kappa$ and $\Delta_{c}$ with respect to the energy scales in atomic dynamics. Then the adiabatic elimination of the cavity mode is applicable and it leads to the relation,
\begin{equation}\label{eq:CA_coupling}
\hat{a}\approx \frac{\eta\int dx \hat{n}(x)\cos\left(k_{0}x\right)}{\tilde{\Delta}_{c}+i\kappa},
\end{equation}
by taking the coarse graining $\delta t_{c}^{-1}\int_{t-\delta t_{c}/2}^{t+\delta t_{c}/2} dt^{'}\hat{A}(t^{'})\approx \hat{A}(t^{'})$ and assuming $\delta t_{c}^{-1}\int_{t-\delta t_{c}/2}^{t+\delta t_{c}/2} dt^{'}\dot{\hat{a}}(t^{'})$ is ignorable in the dynamical equation~\cite{rojan2016localization}. By substituting Eq.~(\ref{eq:CA_coupling}) into Eq. (\ref{eq:Hamiltonian}), we derive
\begin{equation}\label{eq:Hamiltonian_simplified}
\begin{split}
\hat{H}\approx &\int dx \hat{\psi}^{\dagger}(x)[-\frac{\hbar^2\nabla^2}{2m}-\mu]\hat{\psi}(x)\\
&+\eta_{eff}\int dxdx^{'}:\hat{n}(x)\hat{n}(x^{'}):\cos(k_{0}x)\cos(k_{0}x^{'})\\
&+\int\int dx drU(|r|):\hat{n}(x)\hat{n}(x-r):,
\end{split}
\end{equation}
where $\eta_{eff}\approx\eta^{2}\Delta_{c}/(\Delta_{c}^2+\kappa^2)$, after neglecting the Stark shift term by assuming $|\Delta_{c}|\gg V_{0}$ for simplicity. The second term is the cavity-assisted non-local interaction. We would like to note that, in finite-temperature systems, the integration of cavity fields will leave a time-dependent cavity-assisted interaction term~\cite{piazza2013bose}. {While atomic gases generally are trapped with Harmonic trapping potential in experiment, our analysis will be restricted to the ideal case without the trapping potential for simplicity. The focused critical physics is mainly associated with the long-scale degrees of freedom and thus should be insensitive with boundary conditions.}

{The model we consider here is expected to experience a phase transition from the normal phase to superradiant phase when increasing the atom-cavity coupling strength, similar to other intra-cavity quantum gases~\cite{baumann2010dicke,klinder2015observation}. The cavity mode is empty in the normal phase, and is macroscopically occupied in the superradiant phase (i.e., proportional to atom number), respectively~\cite{ritsch2013cold,mivehvar2021cavity}. Here we are mainly interested in the critical behaviours of the superradiant phase transition under the umklapp condition, and thus focus on the low-energy dynamics captured by the bosonization technique, as elaborated below.}

\section{Bosonization}
The single-particle spectra of the Fermi gas have a quadratic dispersion with a Fermi surface determined by the chemical potential $\mu$ at zero temperature [see Fig.~\ref{Fig:model}]. The low-energy dynamics mainly involves the single-particle modes at around the Fermi surface, and then the single-particle part [i.e., the first term in Eq.~(\ref{eq:Hamiltonian})] can be approximately simulated by a linearized Hamiltonian,
 \begin{equation}\label{eq:H0}
\hat{H}_{1}=\sum_{k,s=\pm1}s\upsilon_{f}:\hat{c}_{sk}^{\dagger}\hat{c}_{sk}:=\upsilon_{f}\sum_{s}\int dx :\hat{\psi}_{s}^{\dagger}(-is\partial_{x})\hat{\psi}_{s}:,
\end{equation}
where $\hat{c}_{sk}=L^{-1/2}\int dx \hat{\psi}_{s}e^{is(k+k_{f})x}$ with $s=\pm$ and system length $L$ (also the range of spatial integrals) and $\upsilon_{f}$ is the first-order dispersion coefficient at around the Fermi surface. $\hat{H}_{0}$ can be further bosonized as
\begin{equation}\label{eq:H0_bosonization}
\hat{H}_{1}\approx\frac{\upsilon_{f}}{2}\sum_{s}[\int dx:(\partial_{x}\hat{\phi}_{s})^{2}:+\frac{2\pi}{L}\Delta\hat{N}_{s}(\Delta\hat{N}_{s}+1)],
\end{equation}
with the bosonization relation,
\begin{equation}\label{eq:bosonization}
\hat{\psi}_{\pm}=\frac{\hat{F}_{\pm}}{\sqrt{2\pi\alpha}}e^{\pm2\pi i\Delta \hat{N}_{\pm}x/L}e^{-i\sqrt{2\pi}\hat{\phi}_{\pm}(x)},
\end{equation}
where $\hat{\phi}_{s}$, $\hat{F}_{s}$, $\Delta\hat{N}_{s}$ and $\alpha$ are the Bose field operators, Klein factors, number operators and factor to violate the ultraviolet divergence (the minimum spatial resolution, which in general is set as the lattice constant in lattice model), respectively~\cite{von1998bosonization,gogolin2004bosonization}.

The atom-cavity term [i.e., the second term in Eq.~(\ref{eq:Hamiltonian})] is expanded into,
\begin{equation}\label{eq:HAC_bosonization}
\begin{split}
&\hat{H}_{2}=\frac{\eta_{eff}}{2}\sum_{s_{1}\cdots s_{4}=\pm1}\int dx dr [\cos(2k_{0}x)+\cos(k_{0}r)]\\
&\times:\hat{\psi}_{s_{1}}^{\dagger}(x+\frac{r}{2})\hat{\psi}_{s_{3}}^{\dagger}(x-\frac{r}{2})\hat{\psi}_{s_{4}}(x-\frac{r}{2})\hat{\psi}_{s_{2}}(x+\frac{r}{2}):\\
&\times e^{ik_{f}(s_{2}+s_{4}-s_{1}-s_{3})x+ik_{f}(s_{2}+s_{3}-s_{1}-s_{4})r/2},
\end{split}
\end{equation}
with $\hat{\psi}=\sum_{s}e^{isk_{f}x}\hat{\psi}_{s}$. In order to make the expressions in the summation non-vanishing, $s_{2}+s_{4}-s_{1}-s_{3}$ and $s_{2}+s_{3}-s_{1}-s_{4}$ have to be $\pm4$ and $0$, or $0$ and $\pm4$, which are only satisfied by $(s_{1},s_{2},s_{3},s_{4})=(1,-1,1,-1)$, or $(1,-1,-1,1)$, or $(-1,1,1,-1)$, or $(-1,1,-1,1)$. By employing Eq.~(\ref{eq:bosonization}), we finally derive the bosonized Hamiltonian,
\begin{equation}\label{eq:bosonization_H}
\hat{H}_{2}=\frac{\xi}{L}\int\int dx dr \cos(2\sqrt{\pi}\phi^{(+)}) \cos(2\sqrt{\pi}\phi^{(-)}),
\end{equation}
where the abbreviated expressions $f^{(\pm)}=f(x\pm r/2,\tau)$ are used, $\xi=L\eta_{eff}/(4\pi\alpha)^2$ by considering $\Omega_{c}$ is proportional to the cavity length and further is proportional to the system size in the thermal dynamics . It is necessary to scale the strength of cavity-atom coupling $\xi$ with the system length $L$ to avoid the energy divergence.

The local interaction term [i.e., the last term in Eq.~(\ref{eq:Hamiltonian})] is transformed into
\begin{equation}\label{eq:Hint}
\begin{split}
\hat{H}_{3}=&\sum_{s_{1}\cdots s_{4}=\pm1}\int\int dx drU(r)e^{ik_{f}[(s_{2}+s_{4}-s_{1}-s_{3})x+(s_{3}-s_{4})r]}\\
&\times :\hat{\psi}_{s_{1}}^{\dagger}(x)\hat{\psi}_{s_{3}}^{\dagger}(x-r)\hat{\psi}_{s_{4}}(x-r)\hat{\psi}_{s_{2}}(x):.
\end{split}
\end{equation}
Assuming the characteristic length of $U(|r|)$ is far larger than $2\pi/k_{f}$ but is far smaller than system size, then we can approximately derive
\begin{equation}\label{eq:Hint_bose}
\hat{H}_{3}\approx U_{0}\int dx:(\partial_{x}\phi)^{2}:,
\end{equation}
where $U_{0}$ is the interaction coefficient and $\phi=(\phi_{1}-\phi_{-1})/\sqrt{2}$.

We finally derive the bosonized Hamiltonian,
\begin{equation}\label{eq:bosonization_H}
\begin{split}
\hat{H}=&\sum_{n=1}^{3}\hat{H}_{n}=\frac{ u}{2}\int dx[g(\partial_{x}\theta)^{2}+g^{-1}(\partial_{x}\phi)^{2}]\\
&+\frac{\xi}{L}\int\int dx dr \cos(2\sqrt{\pi}\phi^{(+)}) \cos(2\sqrt{\pi}\phi^{(-)}),
\end{split}
\end{equation}
where $\theta=(\phi_{1}+\phi_{-1})/\sqrt{2}$ and the effective velocity $u=\sqrt{\upsilon_{f}(\upsilon_{f}+2U_{0})}$ and the Luttinger parameter $g=\sqrt{\upsilon_{f}/(\upsilon_{f}+2U_{0})}$. $g>1$ ($g<1$) correspond to the attractive (repulsive) local interactions, respectively.

The Lagrange of our model takes the form,
\begin{equation}\label{eq:La}
\begin{split}
\mathcal{L}=&-i\partial_{x}\theta\partial_{\tau}\phi+\frac{ u}{2}[g(\partial_{x}\theta)^{2}+g^{-1}(\partial_{x}\phi)^{2}]\\
&+\frac{\xi}{L}\int\int dr \cos(2\sqrt{\pi}\phi^{(+)}) \cos(2\sqrt{\pi}\phi^{(-)}),
\end{split}
\end{equation}
with the imaginary time $\tau$ under the convention of action $S=\int \int dxd\tau \mathcal{L}$ and partition function $Z=\int \mathcal{D}[\phi,\theta]e^{-S}$. By integrating the field $\theta$, we yield
\begin{equation}\label{eq:La_phi}
\begin{split}
\mathcal{L}=&\frac{1}{2g}[ u^{-1}(\partial_{\tau}\phi)^{2}+ u(\partial_{x}\phi)^{2}]\\
&+\frac{\xi}{L}\int dr \cos(2\sqrt{\pi}\phi^{(+)}) \cos(2\sqrt{\pi}\phi^{(-)}).
\end{split}
\end{equation}
This is an all-to-all sine-Gordon model in the sense that two fields with any distances are coupled in the cosine term.

\section{Renormalization group analysis}
An (1+1)-dimensional SG model in general has a phase transition at a critical coupling strength, which belongs to the same universality class of KT phase transition in two-dimensional XY model and superfluids, and also characterizes the metal-insulator transition of topological edge states in topological matters~\cite{gogolin2004bosonization,wen2004quantum,cheng2012superconducting}, etc.. The superradiant phase transition is linked to the (1+1)-dimensional KT phase transition of the all-to-all SG model, where the normal or superradiant phases correspond to the irrelevant (gapless) and relevant (gapped) regimes of the cosine term. Inspired by this interesting connection, we will analyze the KT phase transition of the all-to-all SG model in Eq.~(\ref{eq:La_phi}) with perturbative renormalization group technique below.

For discussion convenience, we define $\boldsymbol{x}=(x,-\tau)$ and $\boldsymbol{k}=(k,\omega)$, where $k$ and $\omega$ are the momentum and frequency. To figure out the renormalization process, we set a cutoff $\sqrt{u^{-1}\omega^{2}+u k^{2}}<\Lambda$ on the momentum-frequency plane and denote the field with $\phi_{\Lambda}(\boldsymbol{x})$. The strategy for the perturbative renormalization is standard: 1) expanding the field into low-frequency and high-frequency parts, and integrating out the high-frequency part; 2) rescaling the dimensions back to the same cutoff and then obtaining the renormalization equations.

The field can be expanded into the high and low-frequency components,
\begin{equation}\label{eq:frequency_split}
\phi_{\Lambda}=\frac{1}{\sqrt{\beta L}}(\sum_{|\boldsymbol{k}|<\Lambda^{'}}+\sum_{\Lambda^{'}<|\boldsymbol{k}|<\Lambda})\phi_{\boldsymbol{k}}e^{i\boldsymbol{k}\cdot\boldsymbol{x}}=\phi_{\Lambda^{'}}(\boldsymbol{x})+h(\boldsymbol{x}),
\end{equation}
with $\delta\Lambda=\Lambda-\Lambda^{'}\ll\Lambda$.

First, we need to calculate the effective action for $\phi_{\Lambda^{'}}$ by integrating out $h$ field. As the action $S[h]$ is in the quadratic form, it directly gives $\langle\phi_{\boldsymbol{k}}\phi_{\boldsymbol{k}^{'}}\rangle_{h}=( u^{-1}\omega^{2}+ up^{2})^{-1}\delta_{\boldsymbol{k}^{'},-\boldsymbol{k}}$, and
\begin{equation}\label{eq:h2}
\begin{split}
\langle h^{(\lambda)}h^{(\lambda^{'})}\rangle_{h}&=\frac{g}{\beta L}\sum_{\Lambda^{'}<|\boldsymbol{k}|,|\boldsymbol{k}^{'}|<\Lambda}\frac{\delta_{\boldsymbol{k}^{'},-\boldsymbol{k}}e^{i(\boldsymbol{k}\cdot\boldsymbol{x}^{(\lambda)}+\boldsymbol{k}^{'}\cdot\boldsymbol{x}^{(\lambda^{'})})}}{ u^{-1}\omega^{2}+ up^{2}}\\
&=G_{\lambda\lambda^{'}}dl+\mathcal{O}(dl^{2}),\quad\lambda,\lambda^{'}=\pm,
\end{split}
\end{equation}
where $dl=\ln(\Lambda/\Lambda^{'})$, $G_{++}=G_{--}=g/(2\pi)$ and $G_{+-}=G_{+-}\approx gJ_{0}(\Lambda r/\sqrt{u})/(2\pi)$. Then it leads to
\begin{equation}\label{eq:delta_S_further}
\begin{split}
&\langle\delta S\rangle_{h}=\frac{\xi(1-2gdl)}{L}\int d^{2}\boldsymbol{x} dr\cos(2\sqrt{\pi}\phi_{\Lambda^{'}}^{(+)})\cos(2\sqrt{\pi}\phi_{\Lambda^{'}}^{(-)})\\
&+\frac{2g\xi dl}{L}\int d^{2}\boldsymbol{x} dr J_{0}(\frac{\Lambda r}{\sqrt{u}})\sin(2\sqrt{\pi}\phi_{\Lambda^{'}}^{(+)})\sin(2\sqrt{\pi}\phi_{\Lambda^{'}}^{(-)}).
\end{split}
\end{equation}

\begin{figure}
  \centering
  \includegraphics[width=7.5cm]{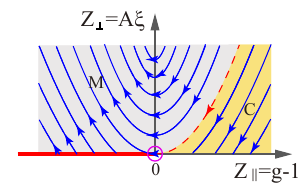}\\
  \caption{Phase diagram. {The bold red line denotes the unstable fixed points on the negative semiaxis of $z_{\perp}$, which corresponds to the quantum phase transition with vanishing coupling strength. The red dashed curve marks the marginal curve of relevant regime, which separates the massive (M) phase (gray shade) and critical (C) phase  (yellow shade). In phase M (C) the system has finite (vanishing) mass gap.} The red circle indicates the nesting effect (infinitely small critical coupling strength) in free Fermi gases predicted with the Landau theory of phase transition~\cite{piazza2014umklapp,keeling2014fermionic,chen2014superradiance}. The phase diagram shows that the nesting effect survives for non-attractive local interactions ($g<1$) only.}\label{Fig:PD}
\end{figure}

Noting that $J_{0}(\Lambda r/\sqrt{u})$ quickly decay with $r$ from $r=0$ as $\Lambda$ is large, we have $2\sin(2\sqrt{\pi}\phi_{\Lambda^{'}}^{(+)})\sin(2\sqrt{\pi}\phi_{\Lambda^{'}}^{(-)})\approx1-2\pi r^{2}(\partial_{x}\phi_{\Lambda^{'}})^{2}-\cos(4\sqrt{\pi}\phi_{\Lambda^{'}})$. Ignoring the irrelevant terms and rescaling the dimensions as $\boldsymbol{k}\rightarrow \Lambda\boldsymbol{k}/\Lambda^{'}$ and $\boldsymbol{x}\rightarrow \Lambda^{'}\boldsymbol{x}/\Lambda$ (note that the system size $L$ is also rescaled), and on the other hand, considering the scaling of $u$ can be absorbed by redefining the coordinates, we yield the following effective lagrangian
\begin{equation}\label{eq:eff_La}
\begin{split}
\mathcal{L}_{eff}=&\frac{1}{2g^{'}}[ u^{-1}(\partial_{\tau}\phi_{\Lambda})^{2}+ u(\partial_{x}\phi_{\Lambda})^{2}]\\
&+\frac{\xi^{'}}{L}\int dr\cos(2\sqrt{\pi}\phi_{\Lambda}^{(+)})\cos(2\sqrt{\pi}\phi_{\Lambda}^{(-)}),
\end{split}
\end{equation}
where $g^{'-1}=[1-2Ag^{2}\xi dl/ L]^{1/2}g^{-1}$, and $\xi^{'}=\xi[1+2(1-g)dl]$, and non-universal coefficient $A=2\pi\int dr r^2 J_{0}(\Lambda r/\sqrt{u})$.
Then the renormalization equations (i.e., the KT equations) are given by
\begin{equation}\label{eq:RGE}
\frac{d\xi}{dl}=2(1-g)\xi,\quad
\frac{dg}{dl}=A\xi g^3,
\end{equation}
Unlike an ordinary SG model~\cite{gogolin2004bosonization}, where the complete RG equations for both the coupling strength and interactions can only obtained with both the one-order and second-order perturbations, the all-to-all SG model is not renormalizable for second-order perturbations.

\section{Phase diagram}
At around the invariant point $g=1$ and $\xi=0$, we define the small fluctuations of parameters
\begin{equation}\label{eq:parameter_fluctuations}
z_{\parallel}=g-1,\quad
z_{\perp}=A\xi.
\end{equation}
The linearized RG equations can be derived in the following forms,
\begin{equation}\label{eq:linearization}
\frac{dz_{\perp}}{dl}=-2z_{\parallel}z_{\perp},\quad
\frac{dz_{\parallel}}{dl}=-z_{\perp},
\end{equation}
which allows us to plot the phase diagram on the plane of $z_{\parallel}$-$z_{\perp}$ like the ordinary local SG model~\cite{gogolin2004bosonization}, as shown in Fig.~\ref{Fig:PD}. The quantity $C=z_{\perp}-z_{\parallel}^{2}$ is invariant along the renormalization flow.

{From Fig.~\ref{Fig:PD}, we can find the RG flows are stationary on the $Z_{\parallel}$ axis. The positive semiaxis of $Z_{\parallel}$ corresponds to stable fixed points and does not indicate any phase transition. The fixed points on the negative semiaxis (see the red bold line) are unstable and correspond to the quantum phase transition toward the relevant regime of cosine perturbation, i.e., the massive (M) phase (see the gray shade), in which $z_{\perp}$ flows to finite along the RG flows. On the critical points, the system is characterized by Gaussian model without the nonlocal cosine term. At around these critical points, the correlation length may be estimated as $\lambda_{C}\propto |\xi|^{1/[2(g-1)]}$ with the critical exponent $1/[2(g-1)]$ by ignoring the variation of $z_{\parallel}$, following the spirit of that the renormalized coupling coefficient $z_{\perp}$ reaches to the order of $1$ when the system size $L\sim \lambda_{C}$ (let $z_{\perp}\sim1$ after integrating the left equation in Eq.~\ref{eq:linearization} with fixed $z_{\parallel}=g-1$)~\cite{gogolin2004bosonization,tzeng2008fidelity}. Then the mass gap is scaled as $E_{g}\propto\lambda_{C}^{-1}\propto |\xi|^{1/[2(1-g)]}$. The critical curve of the unstable fixed point at the origin (see the red dashed curve in Fig.~\ref{Fig:PD}) that separates the M phase and the critical (C) phase~\cite{gogolin2004bosonization} (see yellow shade in Fig.~\ref{Fig:PD}; i.e., the irrelevant regime) corresponds to the KT phase transition with finite cosine term (like finite-temperature KT phase transition in classical models), at around which the system generally has no standard power-law scaling for correlation length.}

{Considering the coupling coefficient of the SG model $\xi\propto\eta_{eff}\propto \eta^{2}$, one can realize that the relevancy of cosine perturbation in the RG analysis is directly linked to the superradiant phase transition of the original intra-cavity Fermi model (\ref{eq:Hamiltonian}). The finite mass gap in phase M corresponds to the band gap opened by the cavity-assisted optical lattice in the superradiant phase.  The continuous symmetry $\phi\rightarrow\phi+\varphi$ of the derivative terms, where $\varphi$ is an arbitrary number, is broken into the discrete symmetry $\phi\rightarrow\phi+n\sqrt{\pi}/2$ with integer $n$ by the presence of cosine term. It corresponds to that the cavity mode is macroscopically occupied and the $Z_{2}$ symmetry of model~(\ref{eq:Hamiltonian}): $\hat{a}\rightarrow -\hat{a}$ and $\hat{\psi}\rightarrow-\hat{\psi}$, is spontaneously broken tn the superradiant phase~\cite{ritsch2013cold,mivehvar2021cavity}. Then we can acquire the knowledge of superradiant phase transition from the above RG analysis of the all-to-all SG model directly. The superradiant phase transition has vanishing critical coupling strengthes for non-attractive local interaction, i.e., when $g\leq 1$. This result is consistent with Refs.~\cite{piazza2014umklapp,keeling2014fermionic,chen2014superradiance}, which predict the nesting effect in the superradiant phase transition of non-interacting degenerate Fermi gases (i.e., $g=1$). The critical exponent of the correlation length is given by $1/[2(g-1)]$, which becomes divergent in the free-fermion limit. The phase boundary for the attractive local interaction $g>1$ is a quadratic curve [see the red dashed curve in Fig.~\ref{Fig:PD}], which is consistent with Refs. ~\cite{chen2015superradiant,yu2018topological} that predict the attractively interacting Fermi gases have finite critical superradiant coupling strength. Noting that $g=\sqrt{\upsilon_{f}/(\upsilon_{f}+2U_{0})}$ and then $(g-1)\approx -U_{0}/\upsilon_{f}$ when $U_{0}/\upsilon_{f}\ll 1$, and $\xi\propto \eta^{2}$, the critical superradiant coupling strength $\eta_{c}\propto |U_{0}|/\upsilon_{f}$, i.e., the critical coupling strength is linearly (inversely) proportional to the local interaction strength $|U_{0}|$.}

{\section{Scaling dimension}}
Now let us go forward to the analysis of scaling dimension of the all-to-all cosine term follow Ref.~\cite{gogolin2004bosonization}. By redefining the fields $\phi/\sqrt{g}\rightarrow\phi$, the Luttinger parameter $g$ will be absorbed into the cosine term:
\begin{equation}\label{eq:La_deformed}
\begin{split}
\mathcal{L}=&\frac{1}{2}[ u^{-1}(\partial_{\tau}\phi)^{2}+ u(\partial_{x}\phi)^{2}]\\
&+\frac{\xi}{L}\int dr \cos(\Gamma\phi^{(+)}) \cos(\Gamma\phi^{(-)}),
\end{split}
\end{equation}
with $\Gamma=2\sqrt{\pi g}$. For the free field $S_{0}=\int \int dx d\tau [ u^{-1}(\partial_{\tau}\phi)^{2}+ u(\partial_{x}\phi)^{2}]/2$, the generating functional of fields $\phi$ is given by
\begin{equation}\label{eq:generate_func}
\begin{split}
Z[h]=&\int \mathcal{D}[\phi(\boldsymbol{x})]e^{-S_{0}-\int dxd\tau h{\boldsymbol{x}}\phi(\boldsymbol{x})}\\
&=Z[0]\exp[\frac{1}{2}\int\int d\boldsymbol{x}d\boldsymbol{x}^{'}h(\boldsymbol{x})G_{0}(\boldsymbol{x},\boldsymbol{x}^{'})h(\boldsymbol{x}^{'})],
\end{split}
\end{equation}
where $G(\boldsymbol{x},\boldsymbol{x}^{'})$ is the Green's function satisfying $-(u^{-1}\partial_{\tau}^{2}+u\partial_{x}^{2})G_{0}(\boldsymbol{x},\boldsymbol{x}^{'})=\delta(\boldsymbol{x}-\boldsymbol{x}^{'})$. With the complex coordinates $z=\tau+ix/u$ and $\bar{z}=\tau-ix/u$, the Green's function takes the form,
\begin{equation}\label{eq:GF}
G_{0}(z,\bar{z})=\frac{1}{4\pi}\ln(\frac{R^{2}}{z\bar{z}+\alpha^{2}}).
\end{equation}
Note that $G(\boldsymbol{x},\boldsymbol{x}^{'})$ only depends on the difference $(\boldsymbol{x}-\boldsymbol{x}^{'})$ denoted by $z$ and $\bar{z}$ in the above equation.

For a particular choice, $h=h_{0}=i\sum_{j=1}^{N}\Gamma_{j}\delta(\boldsymbol{x}-\boldsymbol{x}_{j})$, the generating functional is given by
\begin{equation}\label{eq:correlator_F}
\begin{split}
&\mathcal{F}(1,2,\dots,N)=Z[h_{0}]/Z[0]\\
&=\Pi_{i>j}(\frac{z_{ij}\bar{z}_{ij}}{\alpha^{2}})^{\Gamma_{i}\Gamma_{j}/4\pi}(\frac{R}{\alpha})^{-(\sum_{j}\Gamma_{j})^2/4\pi},
\end{split}
\end{equation}
where $z_{ij}=z_{i}-z_{j}$. To make $\mathcal{F}$ non-vanishing, it requires $\sum_{j}\Gamma_{j}=0$, since $R/\alpha\rightarrow \infty$ in the thermodynamics limit. These basic results all can be found in the Ref.~\cite{gogolin2004bosonization}.

We are mainly interesting the all-to-all cosine term $\hat{B}(\boldsymbol{x})=\frac{\xi}{L}\int dr \cos(\Gamma \hat{\phi}^{(+)})\cos(\Gamma \hat{\phi}^{(-)})$ here. According to the above results, the correlation function of $\hat{B}(\boldsymbol{x})$ can be written as
\begin{equation}\label{eq:correlation_F_B}
\begin{split}
&\langle\hat{B}(\boldsymbol{x})\hat{B}^{\dagger}(\boldsymbol{x}^{'})\rangle=\frac{\xi^{2}}{16 L^2}\sum_{\sigma_{1},\cdots,\sigma_{4}=\pm 1}\int\int dr dr^{'} \\
&\quad\quad\quad\quad\quad\quad\quad\times \langle e^{i\Gamma\sigma_{1}\hat{\phi}^{(+)}}e^{i\Gamma\sigma_{2}\hat{\phi}^{(-)}}e^{i\Gamma\sigma_{3}\hat{\phi}^{(+)'}}e^{i\Gamma\sigma_{4}\hat{\phi}^{(-)'}}\rangle\\
&=\frac{\xi^{2}}{16 L^2}\sum_{\sigma_{1}+\sigma_{2}+\sigma_{3}+\sigma_{4}=0}\int\int dr dr^{'}\Pi_{i>j}(\frac{z_{ij}\bar{z}_{ij}}{\alpha^2})^{\Gamma_{i}\Gamma_{j}/4\pi},
\end{split}
\end{equation}
where we have defined $\hat{\phi}(z_{1,2},\bar{z}_{1,2})=\hat{\phi}^{(\pm)}$, $\hat{\phi}(z_{3,4},\bar{z}_{3,4})=\hat{\phi}^{(\pm)'}=\hat{\phi}(x^{'}\pm\frac{r^{'}}{2},\tau)$, $\Gamma_{j}=\sigma_{j}\Gamma$ and $z_{ij}=z_{i}-z_{j}$.

The condition $\sigma_{1}+\sigma_{2}+\sigma_{3}+\sigma_{4}=0$ requires $\sigma_{j}$s appear in pairs with opposite signs. Then for any sets of $\sigma_{j}$, the six terms in the continued product in Eq.~(\ref{eq:correlation_F_B}) have four terms with powers of $-\Gamma^{2}/4\pi$ and two terms with powers of $\Gamma^{2}/4\pi$. For the scaling transformation $\boldsymbol{x}\rightarrow\lambda \boldsymbol{x}$ and $L\rightarrow\lambda L$, the correlation function $\langle\hat{B}(\boldsymbol{x})\hat{B}^{\dagger}(\boldsymbol{x}^{'})\rangle$ is scaled as $\lambda^{-\Gamma^{2}/\pi}\langle\hat{B}(\boldsymbol{x})\hat{B}^{\dagger}(\boldsymbol{x}^{'})\rangle$.  Therefore, the scaling dimension of the all-to-all cosine term $\hat{B}(\boldsymbol{x})$ is $d=\Gamma^{2}/2\pi=2g$~\cite{gogolin2004bosonization,francesco2012conformal}. {The above RG analysis then further leads to the conclusion that the KT phase transition of all-to-all SG model also subjects to the upper critical dimension $d_{c}=2g_{c}=2$: the perturbation is relevant (irrelevant) when the scaling dimension is lower (higher) than the critical dimension, similar to an ordinary local SG model~\cite{gogolin2004bosonization}, although the all-to-all SG model involves infinitely long-range coupling.}

{We would like to note that, an unperturbed Gaussian quantum field theory with spatial dimension higher than one, as an expected high-dimensional extension of our model, has a total dimension higher than the critical dimension of the KT phase transition, and its phase transition is well characterized with mean-field theory. For higher-dimensional Fermi gas, the critical coupling strengthes for superradiant phase ransition predicted by the mean-field assumption of cavity field are non-vanishing in both the cases with and without attractive interactions~\cite{piazza2014umklapp,keeling2014fermionic,chen2014superradiance,chen2015superradiant}. }

\section{Conclusion}
We theoretically analyze the {superradiant} phase transitions of 1D correlated Fermi gases with cavity-induced umklapp scattering, based on the bosonization and renormalization group techniques. An all-to-all SG model is derived with the bosonization of Fermi fields. The superradiant phase transition is linked to the (1+1)-dimensional KT phase transition of the SG model. The phase diagram given by RG analysis shows that the nesting effect is preserved only with non-attractive interactions. For attractive Fermi gases, the critical coupling strength becomes finite. {The scaling-dimension analysis of the non-local cosine term also subjects to the critical dimension for the KT phase transition, which is two dimension, like that in an ordinary local sine-Gordon model. Our results are consistent with the studies on infinite-range coupling Heisenberg chains~\cite{li2021long} and Ising models~\cite{binney1992theory}, which predicts vanishing critical interactions, as well as the studies on attractively interacting Fermi gases with BCS-BEC crossover~\cite{chen2015superradiant,yu2018topological}, which predicts non-vanishing superradiant critical coupling strengthes. Our analysis is easily extended to the case of hard-core bosons~\cite{rylands2020photon}.}

\emph{Acknowledgements}.--The author thanks Jin Zhang, Yu Chen, Jiangbin Gong, Qingze Guan and Jianwen Jie for the helpful discussions. This work is supported by the National Natural Science Foundation of China (Grant No. 11904228), the Science Specialty Program of Sichuan University (Grand No. 2020SCUNL210) and the National Natural Science Foundation of China (Grant No. 11804221).

\bibliographystyle{apsrev4-2}
\bibliography{ATA_SGM}
\end{document}